\documentclass[preprint,showpacs,prb,amsmath,amssymb]{revtex4}

\usepackage{graphicx}
\usepackage{dcolumn}
\usepackage{bm}

\usepackage{natbib}

\begin{document}

\title{Some Remarks on  One-Dimensional Force-Free Vlasov-Maxwell Equilibria}
\author{M. G. Harrison}
\email{mikeh@mcs.st-and.ac.uk}
\author{T.Neukirch}
\email{thomas@mcs.st-and.ac.uk}
\affiliation{School of Mathematics and Statistics, University of St. Andrews, 
St. Andrews KY16 9SS, United Kingdom}

\begin{abstract}
The conditions for the existence of force-free non-relativistic translationally invariant  one-dimensional (1D) Vlasov-Maxwell (VM) equilibria are investigated using general properties of the 1D VM equilibrium problem.  As has been shown before, the 1D VM equilibrium equations are equivalent to the motion of a pseudo-particle in a conservative pseudo-potential, with the pseudo-potential being proportional to one of the diagonal components of the plasma pressure tensor.  
The basic equations are here derived in a different way to previous work.
Based on this theoretical framework, a necessary condition on the pseudo-potential (plasma pressure) to allow for force-free 1D VM equilibria is formulated. It is shown that linear force-free 1D VM solutions, which so far are the only force-free 1D VM solutions known, correspond to the case where the pseudo-potential is an attractive central potential. A general class of distribution functions leading to central pseudo-potentials is discussed.

\end{abstract}


\maketitle

\section{Introduction}

Plasma equilibria are suitable starting points for investigations of, for example,  plasma instabilities and plasma waves. For collisionless plasmas, the relevant equilibria are self-consistent solutions of the stationary Vlasov-Maxwell (VM) equations (see e.g. Refs. \onlinecite{Krallbook-1973,Schindlerbook}). 

In the present paper we focus exclusively on non-relativistic one-dimensional quasi-neutral VM equilibria with translational symmetry, with the distribution functions depending only on the Hamiltonian and the two canonical momenta corresponding to the invariant directions (in this paper chosen to be the $x$- and $y$-directions).
A large amount of work on translationally invariant 1D VM equilibria of this kind has been done before (e.g. Refs. \onlinecite{Tonks-1959,Grad-1961,Morozov-1961,Harris-1962,Bertotti-1963,Hurley-1963,Nicholson-1963,Sestero-1964,Sestero-1966,Sestero-1967,Lam-1967,Parker-1967,Lerche-1967,Davies-1968,Davies-1969,Alpers-1969,Su-1971,Kan-1972,Channell-1976,Lemaire-1976,Roth-1976,Mynick-1979a,Lee-1979a,Lee-1979b, Greene-1993,Roth-1996,Attico-1999,Mottez-2003,Mottez-2004,Fu-2005,Yoon-2006}), especially on
one-dimensional current sheets and plasma boundary layers, which are of fundamental importance for the structure and stability of plasmas as many plasma activity processes, e.g. magnetic reconnection,\cite{Biskamp-2000,Priest-2000} happen there preferentially.

In the present paper it is not our aim to add to the plethora of 1D VM solutions existing already, but to use some generic properties of the 1D VM equilibrium problem to investigate the conditions for the existence of force-free 1D VM equilibria (see e.g. Ref. \onlinecite{Tassi-2008}).
An obvious property of 1D VM equilibria is that the structure needs to be in force balance. In the quasi-neutral case this means that the sum of the magnetic pressure and one of the diagonal components of the pressure tensor has to be constant. The relevant component of the pressure tensor is the one for the single spatial coordinate upon which the equilibria depend. In this paper we will choose this coordinate to be $z$, so the component of the pressure tensor in the force balance equation will be $P_{zz}$.
In a number of papers it has also been noticed that the translationally invariant 1D VM problem is equivalent to the motion of a pseudo-particle in a conservative pseudo-potential and/or that the force balance for the 1D VM structure is equivalent to pseudo-energy conservation (see e.g. 
Refs. \onlinecite{Grad-1961, Sestero-1966, Lam-1967,Parker-1967,Lerche-1967, Alpers-1969, Su-1971, Kan-1972, Channell-1976, Mynick-1979a, Lee-1979a, Lee-1979b,Greene-1993, Attico-1999}; the pseudo-particle analogy has also recently been used for MHD equilibria in Ref. \onlinecite{Tassi-2008}). 

Directly connected to the pseudo-particle analogy and the related pseudo-energy conservation law (equivalent to force balance of the 1D VM equilbrium) is the special role played by $P_{zz}$. As was first noticed by Grad\cite{Grad-1961} for the case of vanishing electric potential and only one non-vanishing magnetic field and vector potential component, but otherwise arbitrary distribution functions, the derivative of $P_{zz}$ with respect to the non-vanishing component of the vector potential equals (modulo constants) the current density. This was generalized by Bertotti\cite{Bertotti-1963} who included a non-vanishing electric potential and showed that the partial derivative of $P_{zz}$ with respect to the electric potential is proportional to the charge density (see also Refs. \onlinecite{Schindler-1973,Schindlerbook} for the same conclusion for 2D VM equilibria).
Lerche\cite{Lerche-1967} then generalized this to equilibria with two magnetic field, vector potential  and current density components, but for a restricted class of distribution functions. Channell\cite{Channell-1976} investigated the case of vanishing electric potential and a special class of distribution functions for which he showed that the two components of the
current density are, again modulo constant factors, given by the partial derivatives of the particle density with respect to the components of the vector potential. If investigated carefully one can see that the full expressions including the constant factors are again the partial derivatives of $P_{zz}$ with respect to the components of the vector potential. Using the force balance condition, Mynick et al.\cite{Mynick-1979a} then showed that, independently of the distribution function,  the partial derivatives of $P_{zz}$ with respect to the electric potential and the two components of the vector potential are always proportional to the charge density and the components of the current density, respectively. They also showed that this property is maintained under the assumption of quasi-neutrality. The authors then use this property to construct 1D VM equilibria with certain defined properties.
The special role of $P_{zz}$ in the context of 1D VM equilibria has also been emphasized by Attico and Pegoraro\cite{Attico-1999}, again for the case of vanishing electric potential. Similarly to Channell\cite{Channell-1976}, they used this property to construct a number of special distribution functions for 1D VM equilibria. More recently, Mottez\cite{Mottez-2004} gave a detailed discussion of the role of the full pressure tensor (not just of one component) for distribution functions of the same type as discussed by Channell\cite{Channell-1976}, but including the case of non-vanishing electric potential.

The properties of $P_{zz}$ make it a very useful quantity to start any investigation of 1D VM equilibria, since all other quantities such as particle density, charge density and current density can be derived by differentiation. $P_{zz}$ is also equivalent to the pseudo-potential of the analogous pseudo-particle problem and can thus be used to study the properties of 1D VM equilibria qualitatively without the need to solve the equilibrium differential equations.
In particular, the pseudo-particle analogy can be used to formulate conditions on $P_{zz}$ that it has to satisfy to allow the existence of 1D  force-free VM equilibria.\cite{Tassi-2008}
So far only linear force-free 1D VM equilibria are 
known\cite{Sestero-1967, Bobrova-1979, Correa-Restrepo-1993, Bobrova-2001,Bobrova-2003} and the pseudo-particle analogy can also give some insight into the types of distribution functions permitting a linear force-free solution.

The paper is structured as follows. In Sect. \ref{sec:basictheory} we present the basic general theory of quasi-neutral 1D VM equilibria with three constants of motion, rederiving the basic equations given first by Mynick et al.\cite{Mynick-1979a} directly from the definitions of the basic quantities.  In Sect. \ref{sec:forcefree} we discuss general properties of 1D force-free  VM equilibria based on the one-to-one correspondence of the mathematical problem with the motion of a pseudo-particle in a conservative 2D pseudo-potential and we present our conclusions in Sect. \ref{sec:conclusions}.

\section{Basic theory}

\label{sec:basictheory}

We assume that all quantities depend only on $z$ and that the magnetic field has components $B_{x}$ and $B_{y}$. The magnetic field components are written in terms of a vector potential $\mathbf{A}=(A_x,A_y,0)$ where 
\begin{equation}
  B_{x}=-\frac{dA_{y}}{dz},   
  \label{bx}
\end{equation}
\begin{equation}
  B_{y}=\frac{dA_{x}}{dz},  
  \label{bz}
\end{equation}
and  the electric field is the gradient of an  electric potential $\phi$ such that
\begin{equation}
  \mathbf{E}=-\nabla\phi=-\frac{d\phi}{dz}\mathbf{e_{z}} .
  \label{efield}
\end{equation}
In this case $\mathbf{B}$ and $\mathbf{E}$ automatically satisfy the homogenous steady-state Maxwell equations $\nabla \cdot \mathbf{B}=0$ and $\nabla \times \mathbf{E} =\mathbf{0}$.

Due to the symmetries of the system (time independence and spatial independence of $x$ and $y$) the three obvious constants of motion for each particle species are the Hamiltonian or particle energy for each species $s$,
\begin{equation}
H_{s}=\frac{1}{2} m_{s} (v_x^2+v_y^2+v_z^2)+q_{s}\phi,
  \label{hamiltonian}
\end{equation}
the canonical momentum in the $x$-direction, $p_{xs}$,
\begin{equation}
  p_{xs}=m_{s}v_{x}+q_{s}A_{x},
  \label{px}
\end{equation}
and the canonical momentum in the $y$-direction, $p_{ys}$,
\begin{equation}
  p_{ys}=m_{s}v_{y}+q_{s}A_{y},
  \label{py}
\end{equation}
where $m_{s}$ and  $q_{s}$ are the mass and charge of each species. 

All positive functions $f_s $ satisfying the appropriate conditions for existence of the velocity moments and depending only on the constants of motion,
\begin{equation}
  f_{s}=f_{s}(H_{s},p_{xs},p_{ys}),
  \label{equilibriumf}
\end{equation}
solve the steady-state Vlasov equation
\begin{equation}
  \mathbf{v} \cdot \frac{\partial f_{s}}{\partial \mathbf{r}}+ \frac{q_s}{m_s}(\mathbf{E+v\times B}) \cdot \frac{\partial f_{s}}{\partial \mathbf{v}}=0.
  \label{steadystatevlasov}
\end{equation}
We mention that if the same combination of values for the constants of motion allows particle trajectories in several distinct regions of phase space then it is in principle possible to assign different values to $f_s$ in each region\cite{Grad-1961,Mynick-1979a} (for an example for 2D rotationally symmetric VM equilibria see Ref. \onlinecite{Neukirch-1993}). We will, however, not consider this possibility here.

To calculate 1D VM equilibria we  have to solve the remaining inhomogeneous steady-state Maxwell equations, $\nabla\cdot \mathbf{E} = \sigma/\epsilon_{0}$ and 
$\nabla\times\mathbf{B} =\mu_0 \mathbf{j}$. Using equations (\ref{bx}), (\ref{bz}) and (\ref{efield}) these equations reduce to
\begin{eqnarray}
  -\frac{d^{2}\phi}{dz^2}&=&\frac{1}{\epsilon_{0}}\sigma(A_{x},A_{y},\phi), \label{poisson}
\\
  -\frac{d^{2}A_{x}}{dz^2}&=&\mu_{0}j_{x}(A_{x},A_{y},\phi),  \label{amperex}
\\
  -\frac{d^{2}A_{y}}{dz^2}&=&\mu_{0}j_{y}(A_{x},A_{y},\phi), \label{amperey}
\end{eqnarray}
where the source terms are the electric charge density and the current densities in the $x$ and $y$ directions, which are defined as velocity moments of the equilibrium distribution functions, $f_s$, in the following way
\begin{eqnarray}
  \sigma(A_{x},A_{y},\phi)&=&\sum_{s} q_{s}\int_{-\infty}^{\infty} f_{s}(\frac{m_{s}v^{2}}{2}+q_{s}\phi,m_{s}v_{x}+q_{s}A_{x},m_{s}v_{y}+q_{s}A_{y})d^{3}v, 
\label{sigma}\\
  j_{x}(A_{x},A_{y},\phi)& = &\sum_{s} q_{s}\int_{-\infty}^{\infty} v_{x}f_{s}(\frac{m_{s}v^{2}}{2}+q_{s}\phi,m_{s}v_{x}+q_{s}A_{x},m_{s}v_{y}+q_{s}A_{y})d^{3}v,
\label{jx} \\
  j_{y}(A_{x},A_{y},\phi)&=&\sum_{s} q_{s}\int_{-\infty}^{\infty} v_{y}f_{s}(\frac{m_{s}v^{2}}{2}+q_{s}\phi,m_{s}v_{x}+q_{s}A_{x},m_{s}v_{y}+q_{s}A_{y})d^{3}v .
  \label{jy} 
\end{eqnarray}
Here, the dependence of the charge and current densities on the electric and vector potentials has been made visible explicitly. 

Independently of the distribution functions $f_s$, one can show (see Appendix \ref{sec:appendixa}) that the charge and current density components always satisfy the equations
\begin{eqnarray}
  \frac{\partial \sigma}{\partial A_{x}}+\frac{\partial j_{x}}{\partial \phi}&=&0,   \label{force1}
\\
  \frac{\partial \sigma}{\partial A_{y}}+\frac{\partial j_{y}}{\partial \phi}&=&0,
  \label{force2} \\
   \frac{\partial j_x}{\partial A_y}-\frac{\partial j_y}{\partial A_x}&=&0.
   \label{force3} 
\end{eqnarray}
These equations are completely analogous to the relation derived by Schindler and co-workers\cite{Schindler-1973,Schindlerbook} for the case of distribution functions depending only on the Hamiltonian and a single canonical momentum.

As in that case Eqs. (\ref{force1}) - (\ref{force3}) imply the existence of a potential function $P$, where $P$ is given by
\begin{equation}
  P(A_{x},A_{y},\phi)=\sum_{s} \int_{-\infty}^{\infty} m_{s}v_{z}^2f_{s}d^3v, \label{pot}
\end{equation} 
the $P_{zz}$ component of the pressure tensor.  The charge density,  
$\sigma$, and the current densities, $j_{x}$ and $j_{y}$, are given by partial derivatives of the pressure tensor with respect to the electric potential, $\phi$ and the components of the vector potential, $A_x$ and $A_y$, in the following way (for details see Appendix \ref{sec:appendixa}):
\begin{equation}
  \sigma=-\frac{\partial P}{\partial \phi},
  \label{sigphi}
\end{equation}
\begin{equation}
  j_{x}=\frac{\partial P}{\partial A_{x}}  , \label{jcpx}
\end{equation}
\begin{equation}
  j_{y}=\frac{\partial P}{\partial A_{y}}.
  \label{jcpy} 
\end{equation}
Equations (\ref{sigphi}) - (\ref{jcpy}) are equivalent to the force balance of the VM equilibrium.\cite{Mynick-1979a}

Throughout this work we will always be assuming a quasi-neutral plasma consisting of two species (electrons and ions) of opposite charge ($|q_s|=e$). The assumption of  quasineutrality is justified as long as typical length scales of variation of the plasma, $L$, are much larger than the Debye length, $\lambda_D$. This condition is usually satisfied very well in the majority of plasma systems. It should be remarked that quasineutrality does not generally imply that the electric field vanishes.\cite{Schindlerbook, Schindler-2002, Neukirch-1993}

Formally, the quasineutrality condition can be regarded as the lowest order term in an expansion of Poisson's equation for the electric potential in the small parameter 
$\epsilon= \lambda_D/L$. In the lowest order of this expansion,
quasineutrality corresponds to the condition
\begin{equation}
  \sigma(A_{x},A_{y},\phi)=-\frac{\partial P}{\partial \phi}=0.
  \label{qncondition}
\end{equation}
This equation implicitly defines the quasineutral electric potential $\phi_{qn}$ where
\begin{equation}
  \phi_{qn}=\phi_{qn}(A_{x},A_{y}).
  \label{qnepotential}
\end{equation} 
Our potential $P$ is then replaced by its quasineutral version $P_{qn}$, where $P_{qn}$  is a function only of $A_x$ and $A_y$:
\begin{equation}
  P_{qn}(A_x,A_y)=P(A_x,A_y, \phi_{qn}(A_x, A_y)).
  \label{qnpressure}
\end{equation}
It should be noted that due to Eq. (\ref{qncondition}), we have
\begin{equation}
\left(\frac{\partial P_{qn}}{\partial A_x} \right)_{A_y} = \left(\frac{\partial P}{\partial A_x} \right)_{A_y,\phi}+
\left(\frac{\partial \phi_{qn}}{\partial A_x} \right)_{A_y}
\left(\frac{\partial P}{\partial \phi} \right)_{A_x,A_y}
= \left(\frac{\partial P}{\partial A_x} \right)_{A_y,\phi},
\label{dPdAeqn}
\end{equation}
where it is understood that on the right hand side we evaluate $P$ using the quasineutral electric potential and the subscripts indicate explicitly which quantities are being kept constant during the differentiation.
An analogous equations holds for the derivative of $P_{qn}$ with respect to $A_y$.
Thus in Eqs. (\ref{jcpx}) and (\ref{jcpy}) we can replace the derivatives of $P$ by derivatives of
$P_{qn}$ in the quasineutral case.\cite{Mynick-1979a}

Using Eqs. (\ref{jcpx}) and (\ref{jcpy}) we can write Amp\`ere's law as
\begin{eqnarray}
  -\frac{d^{2}A_{x}}{dz^2} &= & \mu_{0}\frac{\partial P_{qn}}{\partial A_{x}} ,\label{ham1}\\
  -\frac{d^{2}A_{y}}{dz^2}&=& \mu_{0}\frac{\partial P_{qn}}{\partial A_{y}} .\label{ham2} 
  \end{eqnarray}
The task of finding 1D quasineutral VM equilibria has been reduced to solving these two coupled second order ordinary differential equations.
  
It has been noticed before for particular distribution functions that the charge and current densities can be written as partial derivatives of a single function of the electric and the vector potential\cite{Kan-1972,Lee-1979a,Lee-1979b} as shown in Eqs. (\ref{sigphi})-(\ref{jcpy}). However, as far as we are aware it was not noticed before that this is a general property of the 1D VM equilibrium problem and that the generating function is identical to $P_{zz}$.

Grad\cite{Grad-1961} seems to have been the first to notice the special role played by $P_{zz}$ for the 1D VM equilibrium problem, but he only investigated the special case of vanishing electric potential and only one non-vanishing component of the magnetic field and vector potential. In that case there is no need to invoke the quasi-neutrality condition and only one of the two equations (\ref{jcpx}) and (\ref{jcpy}) is non-trivial. 

Bertotti\cite{Bertotti-1963} generalized this to include the electric potential and one component of the vector potential and showed that the partial derivatives of $P_{zz}$ give rise to the charge and current density, respectively. He also discussed the assumption of quasi-neutrality for this case.

For the $\phi=0$ case and a set a special assumptions for the distribution functions, 
Lerche\cite{Lerche-1967} was able to relate the derivatives of $P_{zz}$ with respect to $A_x$ and 
$A_y$ to the respective current densities. For $\phi=0$ and similar assumption for the distribution functions, other authors\cite{Parker-1967,Su-1971} also noticed that the current density is related to derivatives of a single function with respect to $A_x$ and $A_y$, but did not relate that function to $P_{zz}$.

Channell\cite{Channell-1976} showed that for $\phi=0$ and distribution functions of the type
\[
f_s=f_{s0} \exp(-\beta_s H_s) g_s(p_{xs}, p_{ys}), 
\]
the current density components are given by the equations,
\begin{eqnarray*}
j_x &= & \left(\frac{1}{\beta_e}+\frac{1}{\beta_i} \right) \frac{\partial N}{\partial A_x}, \\
j_y &= & \left(\frac{1}{\beta_e}+\frac{1}{\beta_i} \right) \frac{\partial N}{\partial A_y}. \\
\end{eqnarray*}
Here the plasma has been assumed to consist of two particle species (electron and ions), $\beta_s = 1/(k_B T_s)$ and $N(A_x,A_y)$ is the particle density of one of the species. Another assumption made by Channell\cite{Channell-1976} is that $N_e(A_x,A_y)$ is the same function of $A_x$ and $A_y$ as $N_i(A_x,A_y)$, which has implications for the distribution functions as well. It is easy to see that under these assumptions we obtain
\[
P_{zz} = \left(\frac{1}{\beta_e}+\frac{1}{\beta_i} \right) N(A_x,A_y),
\]
which is, of course, consistent with Eqs. (\ref{jcpx}) and (\ref{jcpy}).

Using force balance as an argument to show the validity of Eqs. (\ref{sigphi}) - (\ref{jcpy}), 
Mynick et al.\cite{Mynick-1979a} then ultimately derived the general theory presented above. 
Their approach is of course completely equivalent to ours, 
but we showed these equations starting from the velocity 
moments and Eqs. (\ref{force1}) - (\ref{force2}).

Attico and Pegoraro\cite{Attico-1999} also noticed the connection between the current density and the partial derivatives of $P_{zz}$, again for the case $\phi=0$ and for basically the same class of distribution functions used by Channell.\cite{Channell-1976}
In their case the distribution functions are constructed by linear superpositions of distribution functions of the Harris sheet\cite{Harris-1962} type
\[
f_s=f_{s0}\exp(-\beta_s H_s)\int \Phi(u_{xs},u_{ys})\exp[\beta_s (  u_{xs} p_{xs} +u_{ys} p_{ys})] du_{xs} du_{ys}.
\]
The authors point out that for their case Eq. (\ref{force3}) is a necessary condition for the existence of a  single function from which the current density can be derived by differentiation, and show that it is satisfied for their class of distribution functions and that this function is $P_{zz}$ (modulo constant factors).
   
One can see  immediately that Eqs.  (\ref{ham1}) and (\ref{ham2}) are equivalent to the equations of motion of a pseudo-particle with coordinates $A_x$, $A_y$ moving in a conservative 
pseudo-potential $\mu_0 P_{qn}(A_x, \,A_y)$.  The total 
pseudo-energy of the pseudo-particle, $E$, is given by
\begin{equation}
 E=\frac{1}{2}\left(\frac{dA_x}{dz}\right)^{2}+\frac{1}{2}\left(\frac{dA_y}{dz}\right)^{2}+\mu_0 P_{qn}(A_{x},A_{y}) = P_T.
 \label{particleenergy}
\end{equation}
which, on the one hand, is also the Hamiltonian for the particle and, on the other hand, is the equilibrium condition for the 1D VM equilibrium stating that the total pressure, $P_T$, for any 1D VM equilibrium is a constant.  Knowledge of $P_{qn}$  as function of $A_x$ and $A_y$ allows us to predict the nature of the solution using pseudo-energy conservation without solving Eqs. (\ref{ham1}) and (\ref{ham2}) explicitly.  Again, one or several of these properties have been noticed by a large number of authors for special cases (see e.g. Refs.
\onlinecite{Grad-1961,Bertotti-1963,Nicholson-1963,Sestero-1966,Lam-1967,Parker-1967,Lerche-1967,Alpers-1969,Su-1971,Kan-1972,Channell-1976,Lemaire-1976,Roth-1976,Mynick-1979a,Lee-1979a,Lee-1979b,Greene-1993,Attico-1999,Mottez-2003}).

\section{Conditions for Force-free 1D VM Equilibria}

\label{sec:forcefree}

The pseudo-particle analogy can be used to specify the necessary conditions that all force-free solutions of the 1D VM equations have to satisfy. So far only linear force-free 
1D VM solutions are known\cite{Sestero-1967,Channell-1976,Bobrova-1979,Correa-Restrepo-1993,Bobrova-2001} and the analogy could help to answer the question (see e.g. Ref.\onlinecite{Tassi-2008}) whether other, in particular non-linear, force-free solutions exist and how to find them.

The force-free condition 
\begin{equation}
\mathbf{j} \times \mathbf{B} = \mathbf{0},
\label{forcefree1}
\end{equation}
implies that the current density $\mathbf{j} = \nabla\times \mathbf{B} /\mu_0$ 
is parallel to $\mathbf{B}$. In the 1D situation 
we discuss in this paper Eq. (\ref{forcefree1}) can be written as
\begin{equation}
\frac{d}{dz}\left( \frac{B^2}{2\mu_0} \right) = 0.
\label{forcefree2}
\end{equation}
Because the total pressure is always constant for 1D VM equilibria, this implies that for a force-free solution $P_{zz}$ must  also be constant. 

At first sight, this may seem to be at variance with Eqs. (\ref{jcpx}) and (\ref{jcpy}) which clearly imply that the current density is only non-zero if the partial derivatives of $P_{zz}$ with respect to $A_x$ and $A_y$ are non-zero. A closer look, however, reveals that the condition $P_{zz} =$ constant only implies
\begin{equation}
\frac{d P_{zz}}{dz} = \frac{d A_x}{dz} \frac{\partial P_{zz}}{\partial A_x} + 
\frac{d A_y}{dz} \frac{\partial P_{zz}}{\partial A_y} =0 ,
\label{forcefree3}
\end{equation}
for the force-free solution only.  Equation (\ref{forcefree3}), which is, of course, equivalent to Eq. (\ref{forcefree1}), can be satisfied for one solution even if the partial derivatives of $P_{zz}$ are non-zero.
When translated into the pseudo-particle picture, we see that to obtain a pseudo-particle trajectory corresponding to a force-free magnetic field, we need a pseudo-potential ($P_{zz}$) which has at least one equipotential line (contour) that is also a particle trajectory. 
This is a necessary condition for the existence of a 1D force-free VM equilibrium. 

Of course, finding such a potential (pressure) is not yet the complete solution of the problem, but only a first step. For a complete solution we also have to find the distribution functions giving rise to the pressure function (pseudo-potential). In simple cases this may be achievable by using the transform methods presented by Channell\cite{Channell-1976} and Attico and Pegoraro.\cite{Attico-1999} In more general cases, numerical methods as described by Mynick et al.\cite{Mynick-1979a} could be used.

There is, however, a well-known family of pseudo-potentials that satisfies the condition of allowing trajectories which are identical to contours of the pseudo-potential. These are attractive central potentials. We have to restrict ourselves to nonsingular pseudo-potentials because the equivalent pressure must be positive and nonsingular. This rules out, for example, all potentials which are negative powers of the radial coordinate. For central pseudo-potentials we have not only pseudo-energy conservation, but also pseudo-angular momentum conservation. The pseudo-angular momentum is given by
\[
L_{pseudo}=A_x \frac{d A_y}{dz} - A_y\frac{d A_x}{dz} = -\left(A_x B_x + A_y B_y\right),
\]
and is  equal to the negative of the magnetic helicity density.

An example of a distribution function resulting in a central attractive potential is given by\cite{Sestero-1967,Bobrova-1979,Correa-Restrepo-1993,Bobrova-2001}
\begin{equation}
f_s = \frac{n_{0s}}{v_{th,s}^3}\exp(-\beta_s \bar{H}_s),
\label{dfgeneral}
\end{equation}
with
\begin{equation}
\bar{H}_s=H_s +\frac{a_s}{m_s} (p_{xs}^2+p_{ys}^2), 
\label{vquadratic}
\end{equation}
where $a_s$ is a dimensionless constant, which can be related to the temperature anisotropy of the distribution function (see e.g. Ref. \onlinecite{Bobrova-2001}).
We remark that the velocity space integral (\ref{pot}) defining $P$ only exists if $a_s > -1/2$. 
Here $n_{0s}$ is a constant normalizing particle density, $\beta_s = 1/k_B T_s$ is the inverse temperature and $v_{th,s}=(m_s \beta_s)^{-1/2}$ is the thermal velocity. 

For the distribution function (\ref{dfgeneral}) the $zz$-component of the pressure tensor is given by
\begin{equation}
P_{zz} = \sum_s  \frac{1}{\beta_s}\exp(-\beta_s q_s \phi) N_s(A_x,A_y),
\label{pzzquadratic}
\end{equation}
where
\begin{equation}
N_s(A_x,A_y) = \bar{n}_{0s} \exp[-r_{s} (A_x^2 + A_y^2)].
\label{quadraticNs}
\end{equation}
In Eq. (\ref{quadraticNs}) we have used the definitions
\begin{eqnarray}
\bar{n}_{0s} &=& \sqrt{8\pi^3}(1+2a_s)^{-1}
n_{0s},
 \label{n0sbar} \\
r_{s} & =&   \beta_s a_s q_s^2/[m_s(1+2a_s)] . \label{r1s} 
\end{eqnarray}

The charge density is calculated using Eq. (\ref{sigphi}):
\begin{equation}
\sigma = \sum_s q_s \exp(-\beta_s q_s \phi) N_s(A_x,A_y),
\end{equation}
and the quasi-neutrality condition $\sigma=0$ then gives
\begin{equation}
\phi_{qn} = \frac{1}{e(\beta_e + \beta_i)} \ln \left(\frac{N_i}{N_e}\right).
\label{phiqnquadratic}
\end{equation}
One can see immediately that the quasi-neutral electric field will only vanish for a choice of parameters such that $N_e(A_x,A_y)\propto N_i(A_x,A_y)$.

The quasi-neutral $P_{zz}$ is given by
\begin{eqnarray}
P_{zz,qn} &=& \frac{\beta_e + \beta_i}{\beta_e \beta_i} 
N_e^{\beta_i/(\beta_e+\beta_i)} N_i^{\beta_e/(\beta_e+\beta_i)} \nonumber \\
& = &
P_0 \exp[-r_{qn}(A_x^2 + A_y^2)],
\label{pzzquadqn}
\end{eqnarray}
with
\begin{eqnarray}
P_0               & = & \frac{\beta_e + \beta_i}{\beta_e \beta_i} \bar{n}_{0e}^{\beta_i/(\beta_e+\beta_i)}
\bar{n}_{0i}^{\beta_e/(\beta_e+\beta_i)},  \label{defp0} \\
r_{qn} &= &  \frac{\beta_e \beta_i}{\beta_e + \beta_i}  e^2 \left[\frac{ a_e}{m_e(1+2 a_e)} +
 \frac{ a_i}{m_i(1+2 a_i)}\right]   \label{r1qn}.
\end{eqnarray}

The $x$- and $y$-components of the current density can be calculated from Eqs. (\ref{jcpx}) and (\ref{jcpy}), resulting in
\begin{eqnarray}
j_x & = & -2P_0  r_{qn}A_x 
                \exp[-r_{qn}(A_x^2 + A_y^2)], \label{jxquad} \\
j_y & = & -2P_0  r_{qn}A_y 
                \exp(-r_{qn}(A_x^2 +A_y^2)].  \label{jyquad}             
\end{eqnarray}

At this point it is convenient to normalize all quantities. We  normalize the magnetic field to a typical value $B_0$ and the coordinate $z$ to a typical length scale $L$. The components of the vector potential are then normalized by $B_0L$  and the normalized coefficient is defined by
$\bar{r}_{qn} = B_0^2 L^2 r_{1,qn}$. The pressure $P$ is normalized by $B_0^2/\mu_0$ and, finally, the current density components are normalized by $\mu_0 L/B_0$. From now on we assume that all quantities have been normalized in the way just described and we suppress the notation for normalization in what follows.

\begin{figure}
 
 \includegraphics[width=0.8\textwidth]{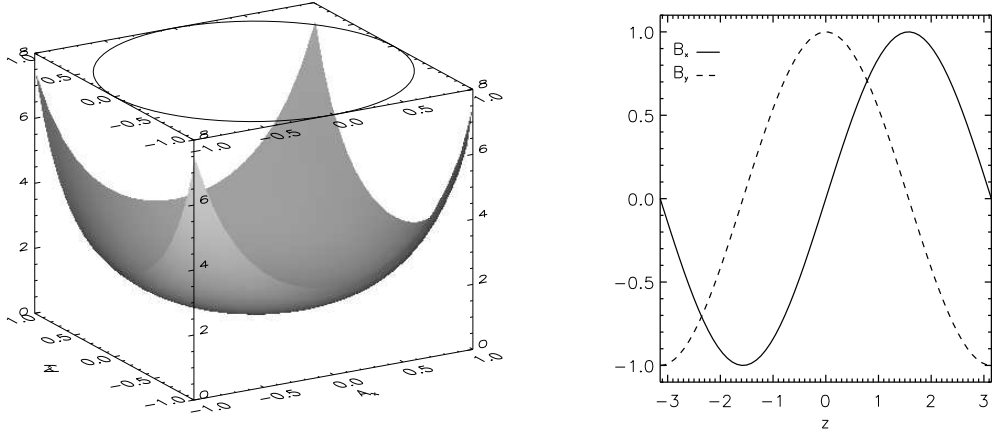}

\caption{Left panel: A surface plot of the $P_{zz}$ for the case when $r_{qn}<0$. The linear force-free solution with $A_x^2+A_y^2 =1$ is shown above the surface plot. Right panel: The magnetic field components as functions of $z$. The magnetic field is normalised to 1 and $\alpha=1$.}
\label{fig:bobrova}
\end{figure}

For the case $r_{qn} < 0$, $P_{zz,qn}$ represents an attractive central pseudo-potential and therefore 
Amp\`ere's law, given by the two coupled differential equations
\begin{eqnarray}
-\frac{d^2 A_x}{dz^2} &=& 2P_0  |r_{qn}|A_x 
                \exp[|r_{qn}|(A_x^2 + A_y^2)],  \label{linearffx} \\
-\frac{d^2 A_y}{dz^2} &=& 2P_0 | r_{qn}|A_y 
                \exp[|r_{qn}|(A_x^2 +A_y^2)],
                 \label{linearffy}
\end{eqnarray}              
allows solutions of the form\cite{Sestero-1967,Bobrova-1979,Correa-Restrepo-1993,Bobrova-2001}
\begin{eqnarray}
A_x & = & k \sin \alpha z , \label{linearffAx} \\
A_y & = & k  \cos \alpha z. \label{linearffAy} 
\end{eqnarray}
These solutions represent circles in the $A_x$-$A_y$-plane and therefore have the property $A_x^2+A_y^2=k^2=$ constant. This renders Eqs. (\ref{linearffx}) and (\ref{linearffy}) linear because the exponential factor is then constant.
For consistency we need that
\begin{equation}
\alpha = \sqrt{2 |r_{qn}| \bar{P}_0 }\exp(|r_{qn}| k^2/2) .
\label{defalpha}
\end{equation}
It is easy to see that 
\begin{eqnarray}
B_x &=&  \alpha A_x = k\alpha \sin \alpha z, \\
B_y &=& \alpha A_y = k\alpha \cos \alpha z,
\end{eqnarray}
so $B_x^2+B_y^2=k^2\alpha^2=$  constant as is required, and that
 \begin{equation}
 \mathbf{j} = \alpha \mathbf {B}.
 \end{equation}
 A surface plot of $P_{zz}$ and the corresponding solution are shown in Fig. \ref{fig:bobrova}. It is obvious that the linear force-free solution is a special solution of the differential equations (\ref{linearffx}) and (\ref{linearffy}) and that other initial conditions will produce non-force-free magnetic fields.
 
Another distribution function giving rise to the same magnetic field solution has been presented by Channell\cite{Channell-1976} (see his case C). In this case a complete solution is available, which has not been discussed in the context of force-free fields, but it is obvious that for the correct choice of initial conditions and parameters the same linear force-free field as given above results. In this case $P_{zz}$ (pseudo-potential) has the form of a 2D harmonic oscillator potential
\begin{equation}
P_{zz}(A_x,A_y) = P_{00} + \frac{1}{2} P_{01} (A_x^2 +A_y^2),
\label{channellffP}
\end{equation}
with $P_{00}$ and $P_{01}$ both positive parameters. 
In this case Amp\`ere's law is always linear and has the form
\begin{eqnarray}
-\frac{d^2 A_x}{dz^2} &=& P_{01}A_x,
                 \label{linearchannellx} \\
-\frac{d^2 A_y}{dz^2} &=& P_{01} A_y.
                 \label{linearchannelly}
\end{eqnarray}            
The general solution as given by Channell is
\begin{eqnarray}
A_x & = & A_{x0} \sin(\sqrt{P_{01}} z + \delta_x), \label{linearaxchannell} \\
A_y & = & A_{y0} \sin(\sqrt{P_{01}} z  + \delta_y). \label{linearaychannell} 
\end{eqnarray}
One can see that by choosing $A_{x0}=A_{y0}=k$, $\delta_x = 0$ and $\delta_y =\pi/2$ we recover the solution from above with $\alpha = \sqrt{P_{01}}$.
The corresponding distribution functions are of the form\cite{Channell-1976}
\begin{equation}
f_s(H_s,p_{xs},p_{ys})  = \exp(-\beta_s H_s) [f_{0s} + f_{1s}(p_{xs}^2+p_{ys}^2)].
\label{channellffdf}
\end{equation}
The same type of $P_{zz}$ and corresponding distribution function has also been found by Attico and Pegoraro,\cite{Attico-1999} but without giving explicit solutions for the magnetic field. 

It is actually straightforward to see that all distribution functions of the type
\begin{equation}
f_s = f_s (H_s, p_s^2),
\label{lffdfgeneral}
\end{equation}
with $p_s^2 = p_{xs}^2+p_{ys}^2$ lead to a $P_{zz}$  corresponding to a central pseudo-potential. Defining $p_{xs} = p\cos \theta$, $p_{ys}=p\sin\theta$, $v_{x} = v \cos \theta$, 
$v_{y}= v \sin\theta$, $A_x = A \cos \theta$, $A_y=A\sin \theta$, so that $v^2= v_x^2+v_y^2$ and $A^2 =A_x^2 + A_y^2$. It is obvious that
\[
p_s^2 = m_s^2 v^2 + 2 m_s q_s A v + q_s^2 A^2,
\]
which does not depend on the angle $\theta$.
Since $H_s$ does not depend upon the vector potential, the integrals defining $P(A_x,A_y)$ (see Eq. (\ref{pot})) define a function depending only upon the magnitude of the vector potential, $A$, but not upon its direction. The corresponding pseudo-potential will therefore be a central potential and if it is attractive it will admit circular orbits, i.e. linear force-free solutions of the same type as discussed before, so obviously there are many distribution functions leading to the same magnetic field solution. 

Another property which is common to all force-free 1D VM solutions is the following. Assume that a 
$P_{zz}(A_x,A_y)$ admitting a force-free solution $A_{x,ff}(z)$ and $A_{y,ff}(z)$ is known, and that the constant value of $P_{zz}$ for the force free solution is $P_{ff}$. Then any (positive) function $F(x)$ can be used to construct a new $\bar{P}_{zz} (A_x,A_y)$ admitting exactly the same force-free solution, by letting
\begin{equation}
\bar{P}_{zz}(A_x,A_y) = \frac{1}{F^\prime(P_{ff})} F(P_{zz}(A_x,A_y)),
\label{newPzz}
\end{equation}
where $F^\prime(x)$ is the derivative of $F$ with respect to its argument.
Using this definition Amp\`ere's law for the new $\bar{P}_{zz}$ has the form
\begin{eqnarray}
-\frac{d^2 A_x}{dz^2} &=& \frac{1}{F^\prime(P_{ff})} F^\prime( P_{zz}(A_x,A_y))
                                                \frac{\partial P_{zz}}{\partial A_x}, \label{newampx}\\
-\frac{d^2 A_y}{dz^2} &=& \frac{1}{F^\prime(P_{ff})} F^\prime( P_{zz}(A_x,A_y))
                                                \frac{\partial P_{zz}}{\partial A_y}. \label{newampy}
\end{eqnarray}
For the force-free solution we have $P_{zz} = P_{ff}$ and thus Eqs. (\ref{newampx}) and (\ref{newampy})
reduce to the equations generated by $P_{zz}$, but only for $(A_x,A_y)=(A_{x,ff},A_{y,ff})$.

\section{Summary and Conclusions}
\label{sec:conclusions}

Previous work\cite{Mynick-1979a} on quasi-neutral 1D VM equilibria has shown that generally the 
quasi-neutral 1D VM equilibrium problem is  equivalent to the equations of motion of a pseudo-particle in a two-dimensional conservative pseudo-potential and that this pseudo-potential is given by one of the diagonal components of the plasma pressure tensor ($P_{zz}$ in the present paper). Furthermore, the partial derivatives of $P_{zz}$ with respect to the electric potential and the two components of the vector potential generate the charge density and the two non-vanishing components of the current density. The equivalence of the 1D VM problem to that of a pseudo-particle moving in a 2D potential is useful because, at least for sufficiently simple potentials, the general nature of the solutions can already be found from a discussion of the pseudo-particle problem using (pseudo-)energy conservation without explicitly solving the equations. 

In the present paper we have first re-derived this general theory directly using the velocity moments and have then applied the pseudo-particle analogy to the interesting problem of  force-free 1D VM solutions, i.e. equilibria where the current density is everywhere parallel or anti-parallel  to the magnetic field. As we were able to show, force-free 1D VM equilibria exist if the pseudo-potential of the equivalent pseudo-particle problem allows a particle trajectory which is identical to a contour of the pseudo-potential. One well-known class of pseudo-potentials allowing this are attractive central potentials, which have circular contours and allow for circular orbits. For these circular orbits, the equivalent 1D VM equilibrium problem is linear in all cases and leads to linear force-free equilibria. We were also able to show that $P_{zz}$ functions corresponding to central potentials can be generated by a whole class of distribution functions with specific properties.
To the best of our knowledge, so far only linear force-free VM equilibria are known\cite{Sestero-1967,Channell-1976, Bobrova-1979,Correa-Restrepo-1993, Bobrova-2001,Bobrova-2003} and it is an interesting question whether nonlinear force-free 1D VM equilibria can be found. We hope that the conditions formulated in this paper will help to find nonlinear force-free VM equilibria.

One motivation for looking for nonlinear force-free 1D VM equilibria is the fact that it is current practice to use pressure balanced VM equilibria with a guide field, mostly the Harris sheet\cite{Harris-1962},  to mimic force-free VM equilibria in, for example, particle-in-cell simulations of magnetic 
reconnection (see e.g. Refs. \onlinecite{Pritchett-2004,Pritchett-2005,Ricci-2004,Hesse-2005,Karimabadi-2005,Silin-2005}). The guide-field equilibria have, however, the same pressure and density gradients as the neutral sheet equilibrium whereas proper force-free equilibria have constant density and pressure. Furthermore, a constant guide field does not add any free energy to the system. Also, unlike for force-free fields, the current density is completely independent of the strength of the added guide field.
It  would therefore be very interesting to investigate whether there are any differences between the reconnection process for guide field systems and for nonlinear force-free systems, for example how micro-instabilities found for systems with plasma density gradients change for force-free equilibria (see e.g. Refs.\onlinecite{Daughton-2003,Silin-2003,Daughton-2004,Ricci-2004b,Karimabadi-2004,Ricci-2005,Silin-2005,Moritaka-2008,Yoon-2008}).

The linear force-free equilibrium 
presented in Section \ref{sec:forcefree} has been investigated using linear\cite{Correa-Restrepo-1993,Bobrova-1979,Bobrova-2001,Bobrova-2003} and 
nonlinear\cite{Bobrova-2001,Bobrova-2003, Li-2003,Bowers-2007} theory, but a detailed comparison of, for example, the magnetic reconnection process with magnetic reconnection in a corresponding neutral sheet has yet to be undertaken. We remark that the doubly periodic nature of this equilibrium may have an effect on the reconnection process compared to, for example, the Harris neutral sheet. It would thus be desirable 
to find a force-free VM equilibrium which has properties similar to the Harris sheet and we believe that the theory presented in this paper is a good point of departure for finding such equilibria.

\acknowledgments{The authors acknowledge support by the UK's Science and Technology Facilities Council.}

\appendix

\section{Derivation of Relations between the $zz$-Component of the Pressure Tensor and the Charge and Current Density}

\label{sec:appendixa}

Following \onlinecite{Schindler-1973,Schindlerbook} the relation (\ref{force1})  can be derived directly by use of the chain rule and by noticing that
\begin{equation}
\int \left(\frac{\partial f_s}{\partial v_x}\right)_{v_y,v_z} d^3v = 0,
\end{equation}
using integration by parts.
This integral vanishes because $f_s$ has to vanish for large $|\mathbf{v}|$.
Also
\begin{equation}
\left(\frac{\partial f_s}{\partial v_x}\right)_{v_y,v_z} = m_s \left[
v_x \left(\frac{\partial f_s}{\partial H_{s}}\right)_{p_{xs},p_{ys}} +
\left(\frac{\partial f_s}{\partial p_{xs}}\right)_{H_s,p_{ys}} \right],
\label{dfsdvxchainrule}
\end{equation}
using $H_s= m_s v^2/2 +q_s \phi$ and $p_{xs} = m_s v_x + q_s A_x$. We obtain
\begin{eqnarray*}
\frac{\partial \sigma}{\partial A_x} + \frac{\partial j_x}{\partial \phi} &=&
\sum_s q_s\left(\frac{\partial }{\partial A_x} \int \, f_s d^3v + \frac{\partial }{\partial \phi} \int \, v_x f_s d^3v\right) \\
&=& \sum_s q_s^2 \int \left[ \left(\frac{\partial f_s}{\partial p_{xs}}\right)_{H_s,p_{ys}} + 
v_x \left(\frac{\partial f_s}{\partial H_s}\right)_{p_{xs},p_{ys}} \right] d^3 v \\
&=& 
\sum_s \frac{q_s^2}{m_s} \int \left(\frac{\partial f_s}{\partial v_x}\right)_{v_y,v_z} d^3v = 0,
\end{eqnarray*}
which leads to Eq. (\ref{force1}). Equation (\ref{force2}) can be derived in exactly the same way by replacing $v_x$ with $v_y$ and $A_x$ by $A_y$.

Equation (\ref{force3}) can be verified by direct differentiation of Eqs. (\ref{jx}) and (\ref{jy}). We see that
\begin{eqnarray*}
\frac{\partial j_x}{\partial A_y} &= &\sum_s q_s^2 \int v_x \left(\frac{\partial f_s}{\partial p_{ys}}\right)_{H_s,p_{xs}} d^3v \\
&=& \sum_s q_s^2 \int \left[\frac{1}{m_s}\left( \frac{\partial (v_x f_s)}{\partial v_y}  \right)_{v_x,v_z} - 
v_x v_y \left(\frac{\partial f_s}{\partial H_s}\right)_{p_{xs},p_{ys}}  \right] d^3v \\
&=& - \sum_s q_s^2 \int v_x v_y \left(\frac{\partial f_s}{\partial H_s}\right)_{p_{xs},p_{ys}} d^3v,
\end{eqnarray*}
where again the chain rule has been used in the first step and integration by parts in the second step. The first term vanishes during integration by parts as any admissible $f_s$ has to go to zero 
as $|\mathbf{v}| \to \infty$ faster than any power of $\mathbf{v}$.
Replacing $A_y$ by $A_x$, $p_{ys}$ by $p_{xs}$ and exchanging $v_x$ and $v_y$ we find with a similar calculation that
\[
\frac{\partial j_y}{\partial A_x} = - \sum_s q_s^2 \int v_x v_y \left(\frac{\partial f_s}{\partial H_s}\right)_{p_{xs},p_{ys}} d^3v,
\]
which shows the general validity of Eq. (\ref{force3}).

Equation (\ref{sigphi}) can also be directly derived by differentiation of the pressure tensor in the following way:
\begin{eqnarray*}
\frac{\partial P}{\partial \phi} &=& \sum_s m_s q_s \int v_z^2 \left( \frac{\partial f_s}{\partial H_s}\right)_{p_{xs},p_{ys}} d^3v \\
&=& \sum_s q_s \int v_z \left(\frac{\partial f_s}{\partial v_z}\right)_{v_x,v_y} d^3 v \\
&=& -\sum_s q_s \int f_s d^3 v = -\sigma,
\end{eqnarray*}
where integration by parts has been used in the last step and 
\begin{equation}
 \left(\frac{\partial f_s}{\partial v_z}\right)_{v_x,v_y} = m_s v_z \left( \frac{\partial f_s}{\partial H_s}\right)_{p_{xs},p_{ys}} 
 \label{dfsdvzchainrule}
 \end{equation}
 in the previous step.
 
Equations (\ref{jcpx}) and (\ref{jcpy}) are similar in structure and thus we show the derivation explicitly only for one of them. Differentiating the $zz$-component of the pressure tensor with respect to $A_x$ we get
 \[
 \frac{\partial P}{\partial A_x} = \sum_s q_s m_s \int  v_z^2 \left(\frac{\partial f_s}{\partial p_{xs}}\right)_{H_s,p_{ys}} d^3v.
 \]
 We can use Eq. (\ref{dfsdvxchainrule}) to express the partial derivative of $f_s$ as
 \[
m_s  \left(\frac{\partial f_s}{\partial p_{xs}}\right)_{H_s,p_{ys}} =
\left(\frac{\partial f_s}{\partial v_x}\right)_{v_y,v_z} - m_s v_x \left( \frac{\partial f_s}{\partial H_s}\right)_{p_{xs},p_{ys}} 
 \]
 and Eq. (\ref{dfsdvzchainrule}) to replace $ m_s v_z\left( \partial f_s/\partial H_s\right)_{p_{xs},p_{ys}} $
 to obtain
 \begin{eqnarray*}
  \frac{\partial P}{\partial A_x} &= & \sum_s q_s \int \left[
  v_z^2\left(\frac{\partial f_s}{\partial v_x}\right)_{v_y,v_z} 
  -v_x v_z \left(\frac{\partial f_s}{\partial v_z}\right)_{v_x,v_y} \right] d^3v \\
  & = & \sum_s q_s \int v_x f_s d^3v = j_x.
 \end{eqnarray*}
We have used integration by parts again in the final step and have retained only the non-vanishing terms. Equation (\ref{jcpy}) follows in exactly the same way by replacing $v_x$ by $v_y$ and $A_x$ by $A_y$.

\bibliographystyle{apsrev}

\end{document}